\pdfoutput=1

\documentclass[12pt,a4paper,final]{article}

\usepackage{ifthen} 
\newboolean{pdflatex}
\setboolean{pdflatex}{true} 

\newboolean{articletitles}
\setboolean{articletitles}{true} 

\newboolean{uprightparticles}
\setboolean{uprightparticles}{false} 

\usepackage{lmodern}
\usepackage{slantsc}

\usepackage{microtype}
\usepackage{lineno}  
\usepackage{xspace} 

\usepackage{graphicx}  
\usepackage{color}
\usepackage{colortbl}
\graphicspath{{./figs/}} 

\usepackage{amsmath} 
\usepackage{amssymb}
\usepackage{amsfonts}
\usepackage{upgreek} 

\usepackage{units}

\usepackage{hyperref}    
\usepackage[all]{hypcap} 

\usepackage[utf8]{luainputenc}
\usepackage{cite} 
\usepackage{mciteplus}

\def\lhcb {\mbox{LHCb}\xspace}
\def\ux85 {\mbox{UX85}\xspace}

\def\babar  {\mbox{BaBar}\xspace}
\def\belle  {\mbox{Belle}\xspace}

\ifthenelse{\boolean{uprightparticles}}%
{

 \def\Pmu         {\ensuremath{\upmu}\xspace}

 \def\Ppi         {\ensuremath{\uppi}\xspace}

 \def\Ppsi        {\ensuremath{\uppsi}\xspace}

 \def\PDelta      {\ensuremath{\Delta}\xspace}                 
 \def\PXi      {\ensuremath{\Xi}\xspace}                 
 \def\PLambda      {\ensuremath{\Lambda}\xspace}                 
 \def\PSigma      {\ensuremath{\Sigma}\xspace}                 
 \def\POmega      {\ensuremath{\Omega}\xspace}                 
 \def\PUpsilon      {\ensuremath{\Upsilon}\xspace}                 
 
 \def\PB      {\ensuremath{\mathrm{B}}\xspace}                 
                  
 \def\PD      {\ensuremath{\mathrm{D}}\xspace}

 \def\PJ      {\ensuremath{\mathrm{J}}\xspace}                 
 \def\PK      {\ensuremath{\mathrm{K}}\xspace}

 \def\Pb      {\ensuremath{\mathrm{b}}\xspace}                 
 \def\Pc      {\ensuremath{\mathrm{c}}\xspace}                 
 \def\Pd      {\ensuremath{\mathrm{d}}\xspace}

 \def\Pi      {\ensuremath{\mathrm{i}}\xspace}

 \def\Pp      {\ensuremath{\mathrm{p}}\xspace}

 \def\Ps      {\ensuremath{\mathrm{s}}\xspace}                 
 \def\Pt      {\ensuremath{\mathrm{t}}\xspace}

}
{

 \def\Pmu         {\ensuremath{\mu}\xspace}

 \def\Ppi         {\ensuremath{\pi}\xspace}

 \def\Ppsi        {\ensuremath{\psi}\xspace}                 
                  
 \mathchardef\PDelta="7101
 \mathchardef\PXi="7104
 \mathchardef\PLambda="7103
 \mathchardef\PSigma="7106
 \mathchardef\POmega="710A
 \mathchardef\PUpsilon="7107
                  
 \def\PB      {\ensuremath{B}\xspace}                 
                  
 \def\PD      {\ensuremath{D}\xspace}

 \def\PJ      {\ensuremath{J}\xspace}                 
 \def\PK      {\ensuremath{K}\xspace}

 \def\Pb      {\ensuremath{b}\xspace}                 
 \def\Pc      {\ensuremath{c}\xspace}                 
 \def\Pd      {\ensuremath{d}\xspace}

 \def\Pi      {\ensuremath{i}\xspace}

 \def\Pp      {\ensuremath{p}\xspace}

 \def\Ps      {\ensuremath{s}\xspace}                 
 \def\Pt      {\ensuremath{t}\xspace}

}

\def\mup        {\ensuremath{\Pmu^+}\xspace}
\def\mun        {\ensuremath{\Pmu^-}\xspace} 

\def\dquark    {\ensuremath{\Pd}\xspace}

\def\squark    {\ensuremath{\Ps}\xspace}

\def\cquark    {\ensuremath{\Pc}\xspace}

\def\bquark    {\ensuremath{\Pb}\xspace}
\def\bquarkbar {\ensuremath{\overline \bquark}\xspace}
\def\bbbar     {\ensuremath{\bquark\bquarkbar}\xspace}
\def\tquark    {\ensuremath{\Pt}\xspace}

\def\pion  {\ensuremath{\Ppi}\xspace}

\def\pip   {\ensuremath{\pion^+}\xspace}
\def\pim   {\ensuremath{\pion^-}\xspace}

\def\kaon  {\ensuremath{\PK}\xspace}
  \def\Kbar  {\kern 0.2em\overline{\kern -0.2em \PK}{}\xspace}

\def\Kz    {\ensuremath{\kaon^0}\xspace}
\def\Kzb   {\ensuremath{\Kbar^0}\xspace}
\def\KzKzb {\ensuremath{\Kz \kern -0.16em \Kzb}\xspace}
\def\Kp    {\ensuremath{\kaon^+}\xspace}
\def\Km    {\ensuremath{\kaon^-}\xspace}

\def\KpKm  {\ensuremath{\Kp \kern -0.16em \Km}\xspace}
\def\KS    {\ensuremath{\kaon^0_{\rm\scriptscriptstyle S}}\xspace} 
\def\KL    {\ensuremath{\kaon^0_{\rm\scriptscriptstyle L}}\xspace} 
\def\Kstarz  {\ensuremath{\kaon^{*0}}\xspace}

  \def\Dbar    {\kern 0.2em\overline{\kern -0.2em \PD}{}\xspace}
\def\D       {\ensuremath{\PD}\xspace}

\def\Dz      {\ensuremath{\D^0}\xspace}
\def\Dzb     {\ensuremath{\Dbar^0}\xspace}
\def\DzDzb   {\ensuremath{\Dz {\kern -0.16em \Dzb}}\xspace}
\def\Dp      {\ensuremath{\D^+}\xspace}
\def\Dm      {\ensuremath{\D^-}\xspace}

\def\DpDm    {\ensuremath{\Dp {\kern -0.16em \Dm}}\xspace}

\def\B       {\ensuremath{\PB}\xspace}
\def\Bbar    {\kern 0.18em\overline{\kern -0.18em \PB}{}\xspace}

\def\Bz      {\ensuremath{\B^0}\xspace}
\def\Bzb     {\ensuremath{\Bbar^0}\xspace}
\def\Bu      {\ensuremath{\B^+}\xspace}
\def\Bub     {\ensuremath{\B^-}\xspace}
\def\Bp      {\ensuremath{\Bu}\xspace}
\def\Bm      {\ensuremath{\Bub}\xspace}

\def\Bd      {\ensuremath{\B^0}\xspace}

\def\Bdb     {\ensuremath{\Bbar^0}\xspace}

\def\jpsi     {\ensuremath{{\PJ\mskip -3mu/\mskip -2mu\Ppsi\mskip 2mu}}\xspace}
\def\psitwos  {\ensuremath{\Ppsi{(2S)}}\xspace}

  \def\Y#1S{\ensuremath{\PUpsilon{(#1S)}}\xspace}

\def\proton      {\ensuremath{\Pp}\xspace}

\def\Lbar {\ensuremath{\kern 0.1em\overline{\kern -0.1em\PLambda}}\xspace}

\newcommand{\decay}[2]{\ensuremath{#1\!\to #2}\xspace}         

\def\to                 {\ensuremath{\rightarrow}\xspace}

\def\CP                {\ensuremath{C\!P}\xspace}

\newcommand{\dmd}{\ensuremath{\Delta m_{\dquark}}\xspace}

\newcommand{\DGd}{\ensuremath{\Delta\Gamma_{\dquark}}\xspace}

\def\AT#1     {\ensuremath{A_{\mathrm{T}}^{#1}}\xspace}           

\def\C#1      {\ensuremath{\mathcal{C}_{#1}}\xspace}                       
\def\Cp#1     {\ensuremath{\mathcal{C}_{#1}^{'}}\xspace}                    
\def\Ceff#1   {\ensuremath{\mathcal{C}_{#1}^{\mathrm{(eff)}}}\xspace}        
\def\Cpeff#1  {\ensuremath{\mathcal{C}_{#1}^{'\mathrm{(eff)}}}\xspace}       
\def\Ope#1    {\ensuremath{\mathcal{O}_{#1}}\xspace}                       
\def\Opep#1   {\ensuremath{\mathcal{O}_{#1}^{'}}\xspace}                    

\newcommand{\tev}{\ensuremath{\mathrm{\,Te\kern -0.1em V}}\xspace}
\newcommand{\gev}{\ensuremath{\mathrm{\,Ge\kern -0.1em V}}\xspace}
\newcommand{\mev}{\ensuremath{\mathrm{\,Me\kern -0.1em V}}\xspace}
\newcommand{\kev}{\ensuremath{\mathrm{\,ke\kern -0.1em V}}\xspace}
\newcommand{\ev}{\ensuremath{\mathrm{\,e\kern -0.1em V}}\xspace}
\newcommand{\gevc}{\ensuremath{{\mathrm{\,Ge\kern -0.1em V\!/}c}}\xspace}
\newcommand{\mevc}{\ensuremath{{\mathrm{\,Me\kern -0.1em V\!/}c}}\xspace}
\newcommand{\gevcc}{\ensuremath{{\mathrm{\,Ge\kern -0.1em V\!/}c^2}}\xspace}
\newcommand{\gevgevcccc}{\ensuremath{{\mathrm{\,Ge\kern -0.1em V^2\!/}c^4}}\xspace}
\newcommand{\mevcc}{\ensuremath{{\mathrm{\,Me\kern -0.1em V\!/}c^2}}\xspace}

\def\mum  {\ensuremath{\,\upmu\rm m}\xspace}

\def\invfb   {\ensuremath{\mbox{\,fb}^{-1}}\xspace}

\def\ps   {\ensuremath{{\rm \,ps}}\xspace}

\def\invps{\ensuremath{{\rm \,ps^{-1}}}\xspace}

\def\gsim{{~\raise.15em\hbox{$>$}\kern-.85em
          \lower.35em\hbox{$\sim$}~}\xspace}
\def\lsim{{~\raise.15em\hbox{$<$}\kern-.85em
          \lower.35em\hbox{$\sim$}~}\xspace}

\def\sPlot{\mbox{\em sPlot}\xspace}

\def\pt         {\mbox{$p_{\rm T}$}\xspace}

\def\evtgen     {\mbox{\textsc{EvtGen}}\xspace}
\def\pythia     {\mbox{\textsc{Pythia}}\xspace}

\def\geant      {\mbox{\textsc{Geant4}}\xspace}

\def\photos     {\mbox{\textsc{Photos}}\xspace}

\def\tell1  {TELL1\xspace}
\def\ukl1   {UKL1\xspace}

\newcommand{\ie}{\mbox{\itshape i.e.}}

\newcommand{\BdToJpsiKS}{\decay{\Bd}{\jpsi\KS}}
\newcommand{\sintwobeta}{\ensuremath{\sin 2 \beta}\xspace}
\newcommand{\SJpsiKS}{\ensuremath{S_{\jpsi\KS }}\xspace}
\newcommand{\CJpsiKS}{\ensuremath{C_{\jpsi\KS }}\xspace}
\def\Kstarzst {\ensuremath{\kaon^{(*)0}}\xspace}

\newcommand*\patchAmsMathEnvironmentForLineno[1]{%
\expandafter\let\csname old#1\expandafter\endcsname\csname
#1\endcsname
\expandafter\let\csname oldend#1\expandafter\endcsname\csname
end#1\endcsname
\renewenvironment{#1}%
{\linenomath\csname old#1\endcsname}%
{\csname oldend#1\endcsname\endlinenomath}}%
\newcommand*\patchBothAmsMathEnvironmentsForLineno[1]{%
\patchAmsMathEnvironmentForLineno{#1}%
\patchAmsMathEnvironmentForLineno{#1*}}%
\AtBeginDocument{%
\patchBothAmsMathEnvironmentsForLineno{equation}%
\patchBothAmsMathEnvironmentsForLineno{align}%
\patchBothAmsMathEnvironmentsForLineno{flalign}%
\patchBothAmsMathEnvironmentsForLineno{alignat}%
\patchBothAmsMathEnvironmentsForLineno{gather}%
\patchBothAmsMathEnvironmentsForLineno{multline}%
}

\usepackage{color}
\usepackage[usenames,dvipsnames,svgnames,table]{xcolor}

\begin{document}

\renewcommand{\thefootnote}{\fnsymbol{footnote}}
\setcounter{footnote}{1}

\begin{titlepage}
\pagenumbering{roman}

\vspace*{-1.5cm}
\centerline{\large EUROPEAN ORGANIZATION FOR NUCLEAR RESEARCH (CERN)}
\vspace*{1.5cm}
\hspace*{-0.5cm}
\begin{tabular*}{\linewidth}{lc@{\extracolsep{\fill}}r}
\ifthenelse{\boolean{pdflatex}}
{\vspace*{-2.7cm}\mbox{\!\!\!\includegraphics[width=.14\textwidth]{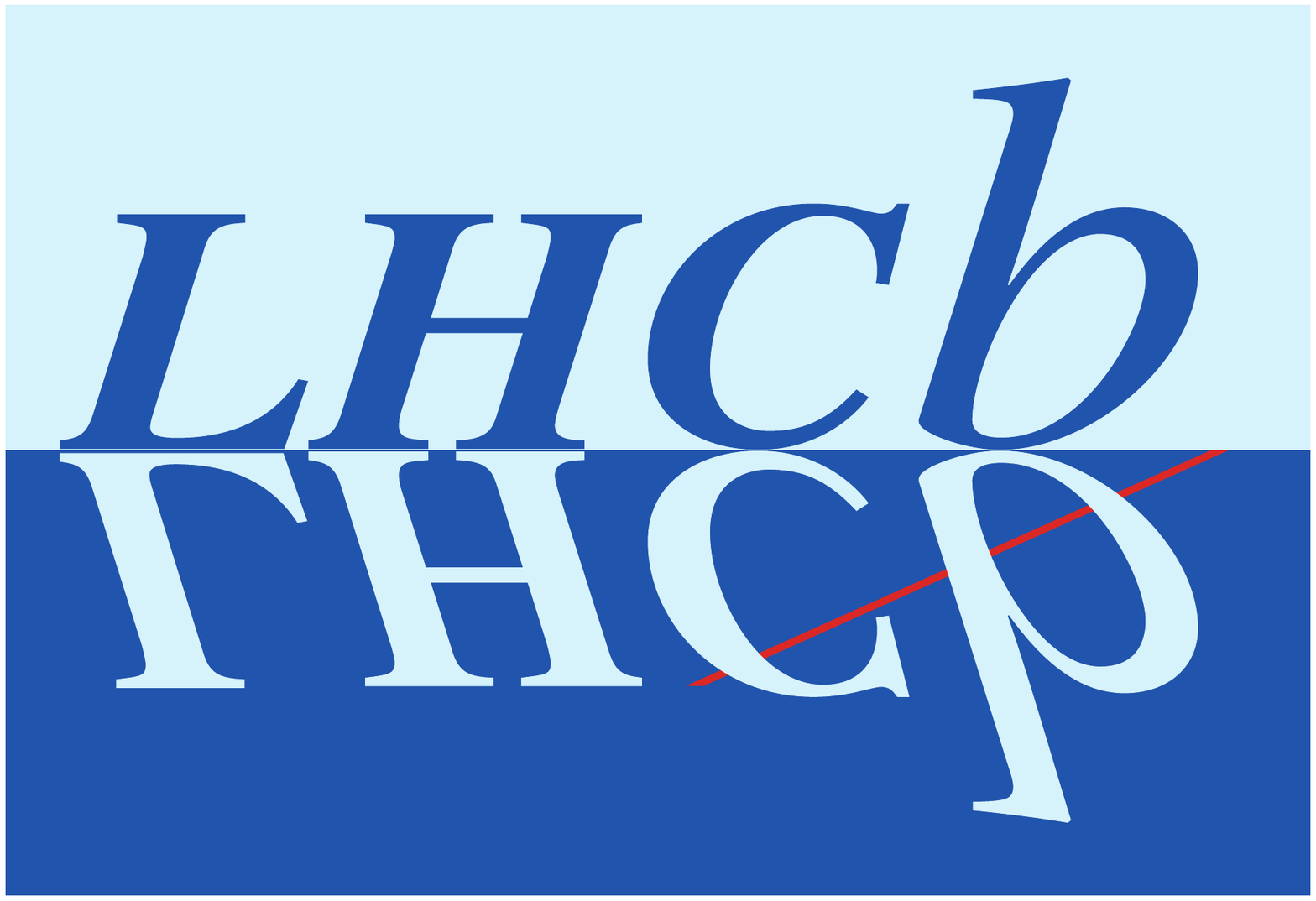}} & &}%
{\vspace*{-1.2cm}\mbox{\!\!\!\includegraphics[width=.12\textwidth]{figs_lhcb-logo.eps}} & &}%
\\
 & & CERN-PH-EP-2012-331 \\  
 & & LHCb-PAPER-2012-035 \\  
 & & November 26, 2012   \\  
 & & \\
\end{tabular*}

\vspace*{4.0cm}

{\bf\boldmath\huge
\begin{center}
  Measurement of the \\time-dependent \CP asymmetry\\ 
  in $\Bd \rightarrow \jpsi \KS$ decays
\end{center}
}

\vspace*{2.0cm}

{\centering 
The LHCb collaboration\footnote{Authors are listed on the following pages.}

}

\vspace{\fill}


\begin{abstract}
\noindent

This Letter reports a measurement of the \CP violation observables \SJpsiKS and
\CJpsiKS in the decay channel $\Bd\to\jpsi\KS$ performed with 1.0\invfb of
$\proton\proton$ collisions at $\sqrt{s}=7\tev$ collected by the LHCb
experiment. The fit to the data yields $\SJpsiKS = 0.73 \pm
0.07\,\text{(stat)}\pm 0.04\,\text{(syst)}$ and $\CJpsiKS = 0.03 \pm
0.09\,\text{(stat)} \pm 0.01\,\text{(syst)}$. Both values are consistent with
the current world averages and within expectations from the Standard Model.

\end{abstract}

%
%
%

\vspace*{2.0cm}
{\centering 
Published as \href{http://dx.doi.org/10.1016/j.physletb.2013.02.054}{Phys. Lett. \textbf{B721} (2013) 24--31}

}

\vspace{\fill}

\end{titlepage}


\newpage
\setcounter{page}{2}
\mbox{~}
\newpage

\centerline{\large\bf LHCb collaboration}
\begin{flushleft}
\small
R.~Aaij$^{38}$, 
C.~Abellan~Beteta$^{33,n}$, 
A.~Adametz$^{11}$, 
B.~Adeva$^{34}$, 
M.~Adinolfi$^{43}$, 
C.~Adrover$^{6}$, 
A.~Affolder$^{49}$, 
Z.~Ajaltouni$^{5}$, 
J.~Albrecht$^{35}$, 
F.~Alessio$^{35}$
% ...

{\footnotesize \it
% ...
$ ^{5}$Clermont Universit\'{e}, Universit\'{e} Blaise Pascal, CNRS/IN2P3, LPC, Clermont-Ferrand, France\\
$ ^{6}$CPPM, Aix-Marseille Universit\'{e}, CNRS/IN2P3, Marseille, France\\
% ...
$ ^{11}$Physikalisches Institut, Ruprecht-Karls-Universit\"{a}t Heidelberg, Heidelberg, Germany\\
% ...
$ ^{33}$Universitat de Barcelona, Barcelona, Spain\\
$ ^{34}$Universidad de Santiago de Compostela, Santiago de Compostela, Spain\\
$ ^{35}$European Organization for Nuclear Research (CERN), Geneva, Switzerland\\
% ...
$ ^{38}$Nikhef National Institute for Subatomic Physics, Amsterdam, The Netherlands\\
% ...
$ ^{43}$H.H. Wills Physics Laboratory, University of Bristol, Bristol, United Kingdom\\
% ...
$ ^{49}$Oliver Lodge Laboratory, University of Liverpool, Liverpool, United Kingdom\\
% ...
$ ^{n}$LIFAELS, La Salle, Universitat Ramon Llull, Barcelona, Spain\\
% ...
}
\end{flushleft}

\cleardoublepage

\renewcommand{\thefootnote}{\arabic{footnote}}
\setcounter{footnote}{0}


\pagestyle{plain} 
\setcounter{page}{1}
\pagenumbering{arabic}

\section{Introduction}
\label{sec:Introduction}

The source of \CP violation in the electroweak sector of the Standard Model (SM)
is the single irreducible complex phase of the Cabibbo-Kobayashi-Maskawa (CKM)
quark mixing matrix~\cite{Kobayashi:1973fv,Cabibbo:1963yz}. The decay
\BdToJpsiKS is one of the theoretically cleanest modes for the study of \CP
violation in the \Bd meson system. Here, the \Bd and \Bdb mesons decay to a
common $\CP$-odd eigenstate allowing for interference through \Bz--\Bzb mixing.

In the \Bd system the decay width difference \DGd between the heavy and light
mass eigenstates is negligible. Therefore, the time-dependent decay rate
asymmetry can be written as~\cite{PhysRevD.23.1567,Bigi198741}
\begin{align}
{\cal A}_{\jpsi\KS}(t) & \equiv
  \frac{\Gamma(\Bdb(t)\to\jpsi\KS) - \Gamma(\Bd(t)\to\jpsi\KS)}
       {\Gamma(\Bdb(t)\to\jpsi\KS) + \Gamma(\Bd(t)\to\jpsi\KS)} \nonumber\\
 &= \SJpsiKS\sin(\dmd t) - \CJpsiKS\cos(\dmd t).
\end{align}
Here $\Bd(t)$ and $\Bdb(t)$ are the states into which particles produced at
$t=0$ as \Bd and \Bdb respectively have evolved, when decaying at time $t$. The
parameter $\dmd$ is the mass difference between the two \Bd mass eigenstates.
The sine term results from the interference between direct decay and decay after
\Bz--\Bzb mixing. The cosine term arises either from the interference between
decay amplitudes with different weak and strong phases (direct \CP violation) or
from \CP violation in \Bz--\Bzb mixing.

In the SM, \CP violation in mixing and direct \CP violation are both negligible
in \BdToJpsiKS decays, hence $\CJpsiKS\approx 0$, while $\SJpsiKS \approx
\sintwobeta$, where the CKM angle $\beta$ can be expressed in terms of the CKM
matrix elements as $\arg\left|-V_{\cquark\dquark}^{\phantom{\ast}}
V_{\cquark\bquark}^{\ast}/V_{\tquark\dquark}^{\phantom{\ast}}
V_{\tquark\bquark}^{\ast}\right|$. It can also be measured in other \Bz decays
to final states including charmonium such as $\jpsi\KL$, $\jpsi\Kstarz$,
$\psitwos\Kstarzst$, which have been used in measurements by the \babar and
\belle collaborations~\cite{Aubert:2009aw,Adachi:2012et}. Currently, the world
averages are $\SJpsiKS = 0.679 \pm 0.020$ and $\CJpsiKS = 0.005 \pm
0.017$~\cite{Amhis:2012bh}.

The time-dependent measurement of the \CP parameters \SJpsiKS and \CJpsiKS
requires flavour tagging, \ie\ the knowledge whether the decaying particle was
produced as a \Bz or a \Bzb meson. If a fraction $\omega$ of candidates is
tagged incorrectly, the accessible time-dependent asymmetry ${\cal
A}_{\jpsi\KS}(t)$ is diluted by a factor $(1-2\omega)$. Hence, a measurement of
the \CP parameters requires precise knowledge of the wrong tag fraction.
Additionally, the asymmetry between the production rates of \Bz and \Bzb has to
be determined as it affects the observed asymmetries.

In this Letter, the most precise measurement of \SJpsiKS and \CJpsiKS to date
at a hadron collider is presented using approximately 8200 flavour-tagged
\BdToJpsiKS decays.

\section{Data samples and selection requirements}
\label{sec:Detector}

The data sample consists of $1.0$\invfb of $\proton\proton$ collisions
recorded in 2011 at a centre-of-mass energy of $\sqrt{s}=7$\tev with the \lhcb
experiment at CERN. The detector~\cite{external:detector} is a single-arm
forward spectrometer covering the \mbox{pseudorapidity} range $2$ to $5$,
designed for the study of particles containing \bquark or \cquark quarks. It
includes a high precision tracking system consisting of a silicon-strip vertex
detector surrounding the $pp$ interaction region, a large-area silicon-strip
detector located upstream of a dipole magnet with a bending power of about
$4$\,Tm, and three stations of silicon-strip detectors and straw drift-tubes
placed downstream. The combined tracking system has a momentum resolution
$\Delta p/p$ that varies from $0.4$\,\% at $5$\gevc to $0.6$\,\% at 100\gevc,
and an impact parameter resolution of $20$\mum for tracks with high transverse
momentum. Charged hadrons are identified using two ring-imaging Cherenkov
detectors. Photon, electron and hadron candidates are identified by a
calorimeter system consisting of scintillating-pad and preshower detectors, an
electromagnetic and a hadronic calorimeter. Muons are identified by a system
composed of alternating layers of iron and multiwire proportional chambers.

The analysis is performed on events with reconstructed $\Bz\to\jpsi\KS$
candidates with subsequent $\jpsi\to\mup\mun$ and $\KS\to\pip\pim$ decays.
Events are selected by the trigger consisting of hardware and software stages.
The hardware stage accepts events if muon or hadron candidates with high
transverse momentum (\pt) with respect to the beam axis are detected. In the
software stage, events are required to contain two oppositely-charged particles,
both compatible with a muon hypothesis, that form an invariant mass greater than
$2.7\gevcc$. The resulting \jpsi candidate has to be clearly separated
(decay length significance greater than $3$) from the
production vertex (PV) with which it is associated on the basis of the impact
parameter. The overall signal efficiency of these triggers is found
to be $64\%$.

Further selection criteria are applied offline to decrease the number of
background candidates. The \jpsi candidates are reconstructed from two
oppositely-charged, well identified muons with $\pt > 500\mevc$ that form a
common vertex with a fit $\chi^2/\text{ndf}$ of less than $11$, where
$\text{ndf}$ is the number of degrees of freedom, and with an invariant mass in
the range $3035$--$3160$\mevcc. It is required that the \jpsi candidate fulfils
the trigger requirements described above. The \KS candidates are formed from two
oppositely-charged pions, both with (long \KS candidate) or without (downstream
\KS candidate) hits in the vertex detector. Any \KS candidates where both pion
tracks have hits in the tracking stations but only one has additional hits in
the vertex detector are ignored, as they would only contribute to $<2\%$ of the
events. Each pion must have $p > 2\gevc$ and a clear separation from any PV.
Furthermore, they must form a common vertex with a fit $\chi^2/\text{ndf}$ of
less than $20$ and an invariant mass within the range $485.6$--$509.6$\mevcc
(long \KS candidates) or $476.6$--$518.6$\mevcc (downstream \KS candidates).
Different mass windows are chosen to account for different mass resolutions for
long and downstream \KS candidates. The \KS candidate's decay vertex is required
to be significantly displaced with respect to the associated PV.

The \Bz candidates are constructed from combinations of \jpsi and \KS candidates
that form a vertex with a reconstructed mass $m_{\jpsi\KS}$ in the range
$5230$--$5330$\mevcc. The value of $m_{\jpsi\KS}$ is computed constraining the
invariant masses of the $\mup\mun$ and $\pip\pim$ to the known \jpsi and \KS
masses \cite{external:PDG}, respectively. As most events involve more than one
reconstructed PV, \Bz candidates are required to be associated to one 
PV only and are therefore omitted if their impact parameter significance with
respect to other PVs in the event is too small. Additionally, the \KS 
candidate's decay vertex is required to be separated from the \Bz decay vertex 
by a decay time significance of the \KS greater than 5.

The decay time $t$ of the $\Bz$ candidates is determined from a vertex fit to
the whole decay chain under the constraint that the \Bz candidate originates
from the associated PV \cite{Hulsbergen2005566}. Only candidates with a good
quality vertex fit and with $0.3<t<18.3$\ps are retained. In case more than one
candidate is selected in an event, that with the best vertex fit quality is
chosen. The fit uncertainty on $t$ is used as an estimate of the decay time
resolution $\sigma_{t}$, which is required to be less than $0.2$\ps. Finally,
candidates are only retained if the flavour tagging algorithms provide a
prediction for the production flavour of the candidate, as discussed in
Section~\ref{sec:tagging}.

Simulated samples are used for cross-checks and studies of decay time
distributions. For the simulation, $pp$ collisions are generated using
\pythia~6.4~\cite{Sjostrand:2006za} with a specific \lhcb
configuration~\cite{LHCb-PROC-2010-056}. Decays of hadronic particles are
described by \evtgen~\cite{Lange:2001uf} in which final state radiation is
generated using \photos~\cite{Golonka:2005pn}. The interaction of the generated
particles with the detector is implemented using the \geant
toolkit~\cite{Allison:2006ve, *Agostinelli:2002hh} as described in
Ref.~\cite{LHCb-PROC-2011-006}.

\section{Flavour tagging}
\label{sec:tagging}

A mandatory step for the study of \CP violating quantities is to tag the
initial, \ie\ production, flavour of the decaying \Bz meson. Since \bquark
quarks are predominantly produced in \bbbar pairs in LHCb, the flavour tagging
algorithms used in this analysis~\cite{internal:tagging} reconstruct the flavour
of the non-signal \bquark hadron. The flavour of the non-signal \bquark hadron
is determined by identifying the charge of its decay products, such as that of
an electron or a muon from a semileptonic \bquark decay, a kaon from a
$\bquark\to\cquark\to\squark$ decay chain, or the charge of its inclusively
reconstructed decay vertex. The algorithms use this information to provide a tag
$d$ that takes the value $+1$ ($-1$) in the case where the signal candidate is
tagged as an initial \Bz (\Bzb) meson.

A careful study of the fraction of candidates that are wrongly tagged (mistag
fraction) is necessary as the measured asymmetry is diluted due to the imperfect
tagging performance. The mistag fraction ($\omega$) is extracted on an
event-by-event basis from the combined per-event mistag probability prediction
$\eta$ of the tagging algorithms. On average, the mistag fraction is found to
depend linearly on $\eta$ and is parameterised as
\begin{align}
\omega(\eta) = p_1\cdot\left(\eta-\langle\eta\rangle\right)+p_0 \ .
\end{align}
Using events from the self-tagging control channel $\decay{\Bu}{\jpsi\Kp}$, the
parameters are determined to be $p_1 = 1.035 \pm 0.021 \,\text{(stat)} \pm 0.012
\,\text{(syst)}$, $p_0 = 0.392 \pm 0.002 \,\text{(stat)} \pm 0.009
\,\text{(syst)}$ and $\langle\eta\rangle = 0.391$~\cite{LHCb-CONF-2012-026}. The
systematic uncertainties on the tagging calibration parameters are estimated by
comparing the tagging performance obtained in different decay channels such as
$\decay{\Bd}{\jpsi\Kstarz}$, in $\Bp$ and $\Bm$ subsamples separately, and in
different data taking periods.

The difference in tagging response between \Bz and \Bzb is parameterised by
using
\begin{align}
  \omega = \omega(\eta) \pm \frac{\Delta p_0}{2} \ ,
\end{align}
where the $+$ ($-$) is used for a $\Bz$ ($\Bzb$) meson at production and $\Delta
p_0$ is the mistag fraction asymmetry parameter, which is the difference of
$p_0$ for $\Bz$ and $\Bzb$ mesons. It is measured as $\Delta p_0 = 0.011 \pm
0.003$ using events from the control channel $\decay{\Bu}{\jpsi\Kp}$. By using
$\Delta p_0$ in the analysis, the systematic uncertainty on the $p_0$ parameter
is reduced to $0.008$. The difference of tagging efficiency for $\Bz$ 
and $\Bzb$ mesons is measured in the same control channel as 
$\Delta \varepsilon_{\text{tag}} = 0.000 \pm 0.001$ and is therefore negligible. 
Thus, it is only used to estimate possible systematic uncertainties in 
the analysis.

The effect of imperfect tagging is the reduction of the statistical power by a
factor $\varepsilon_\text{tag}\mathcal{D}^2$, where $\varepsilon_\text{tag}$ is
the tagging efficiency and $\mathcal{D}=1-2\omega$ is the dilution factor. The
effective $\varepsilon_\text{tag}$ and $\mathcal{D}$ values are measured as
$\varepsilon_\text{tag} = (32.65 \pm 0.31 )\%$ and $\mathcal{D}= 0.270 \pm
0.015$, resulting in $\varepsilon_\text{tag}\mathcal{D}^2=(2.38 \pm 0.27)\%$ ,
where combined systematic and statistical uncertainties are quoted. The
measured dilution corresponds to a mistag fraction of $\omega = 0.365 \pm
0.008$.

\section{Decay time acceptance and resolution}
\label{sec:resolution}

The bias on the decay time distribution due to the trigger is estimated by
comparing candidates selected using different trigger requirements. In the
selection, the reconstructed decay times of the \BdToJpsiKS candidates are
required to be greater than $0.3$\ps. This requirement makes the acceptance
effects of the trigger nearly negligible. However, some small efficiency loss
remains for small decay times. Neglecting this efficiency loss is treated as a
source of systematic uncertainty.

A decrease of efficiency is also observed at large decay times, mostly affecting
the candidates in the long \KS subsample. This can be described with a linear
efficiency function with parameters determined from simulated data for the
downstream and long \KS subsamples separately. The efficiency function is then
used to correct the description of the decay time distribution.

The finite decay time resolution of the detector leads to an additional dilution 
of the experimentally accessible asymmetry. It is modelled 
event-by-event with a triple Gaussian function,
\begin{align}
  \mathcal{R}(t-t'|\sigma_{t}) = \sum_{i=1}^{3} f_{i} 
  \frac{1}{\sqrt{2\pi}s_{i}\sigma_{t}} \exp{\left( - \frac{(t-t'-b\sigma_{t})^2}{2(s_i\sigma_{t})^2} \right)} \ ,
\end{align}
where $t$ is the reconstructed decay time, $t'$ is the true decay time, and
$\sigma_{t}$ is the per-event decay time resolution estimate. The parameters
are: the three fractions $f_i$, which sum to unity, the three scale factors $s_i$,
and a relative bias $b$, which is found to be small. They are determined from a
fit to the $t$ and $\sigma_{t}$ distributions of prompt \jpsi events that pass
the selection and trigger criteria for \BdToJpsiKS, except for decay time
biasing requirements. The parameters are determined separately for the
subsamples formed from downstream and long \KS candidates. This results in an
average effective decay time resolution of $\unit[55.6]{fs}$ ($\unit[65.6]{fs}$)
for candidates with long (downstream) \KS.

\section{\texorpdfstring{Measurement of {\boldmath \SJpsiKS and \CJpsiKS}}{Measurement of S and C}}
\label{sec:fitmodel}

The analysis is performed using the following set of observables: the
reconstructed mass $m_{\jpsi\KS}$, the decay time $t$, the estimated decay time
resolution $\sigma_{t}$, the flavour tag $d$, and the per-event mistag
probability $\eta$. The \CP observables \SJpsiKS and \CJpsiKS are determined as
parameters in an unbinned extended maximum likelihood fit to the data.

Due to different resolution and acceptance effects for the downstream and long
\KS subsamples, a simultaneous fit to both subsamples is performed. In each
subsample, the probability density function (PDF) is defined as the sum of two
individual PDFs, one for each of the components of the fit: the \Bz signal and
the background. The latter component contains both combinatorial background and
mis-reconstructed $b$-hadron decays.

The reconstructed mass distribution of the signal is described by the sum of two
Gaussian PDFs with common mean but different widths. Only the mean is shared 
between the two subsamples. The background component is parameterised as an
exponential function, different for each subsample.

The signal and background distributions of the per-event mistag probability
$\eta$ are modelled with PDFs formed from histograms obtained with the \sPlot
technique \cite{external:splot} on the reconstructed mass distribution. In both
subsamples the same signal and background models are used.

The distributions of the estimated decay time resolution $\sigma_{t}$ are
different in each component and each subsample. Hence, no parameters are shared
between subsamples or components. All $\sigma_{t}$ PDFs are modelled with lognormal
functions
\begin{align}
  \text{Ln}(\sigma_{t};M_{\sigma_{t}},k) 
    = \frac{1}{\sqrt{2\pi} \sigma_{t} \ln k} 
      \exp\left( - \frac{\ln^2(\sigma_{t}/M_{\sigma_{t}})}{2 \ln^2(k)} \right) ,
\end{align}
where $M_{\sigma_{t}}$ is the median and $k$ the tail parameter. The
background components in both subsamples are parameterised by single lognormal
functions. For the signal a sum of two lognormals with common (different) median 
parameter(s) is chosen for the long \KS (downstream \KS) subsample. 

The background PDFs of the decay time are modelled in each subsample by the
sum of two exponential functions. These are convolved with the corresponding
resolution function $\mathcal{R}(t-t'|\sigma_{t})$. The parameters are not shared 
between the two subsamples. The background distribution
of tags $d$ is described as a uniform distribution.

The signal PDF for the decay time simultaneously describes the
distribution of tags $d$, and is given by
\begin{align}
  \mathcal{P}(t,d|\sigma_{t},\eta) = \epsilon(t) \cdot \mathcal{P}_{\CP}(t',d|\sigma_{t},\eta) \otimes \mathcal{R}(t-t'|\sigma_{t}) \ , 
\end{align}
with
\begin{align}
  \mathcal{P}_{\CP}(t',d|\sigma_{t},\eta) \propto 
  e^{-t'/\tau} \Big( 
                1 & - d \Delta p_0 - d A_{\text{P}}  (1-2\omega(\eta)) \notag \\
                  & - (d (1-2\omega(\eta)) - A_{\text{P}}  (1-d\Delta p_0)) \SJpsiKS \sin \dmd t' \notag \\
                  & + (d (1-2\omega(\eta)) - A_{\text{P}}  (1-d\Delta p_0)) \CJpsiKS \cos \dmd t'
              \Big) \ .
\end{align}
This PDF description exploits time-dependent asymmetries, while its
normalisation adds sensitivity by accessing time-integrated asymmetries. The
lifetime $\tau$, the mass difference \dmd, and the \CP parameters \SJpsiKS and
\CJpsiKS are shared in the PDFs of the downstream and long \KS subsamples, as
well as the asymmetry $A_{\text{P}} = (R_{\Bzb} - R_{\Bz})/(R_{\Bzb} + R_{\Bz})$
of the production rates $R$ for \Bzb and \Bz mesons in $pp$ collisions at LHCb.
The latter value has been measured in
Refs.~\cite{LHCb-PAPER-2011-029,LHCb-CONF-2012-007} to be $A_{\text{P}}
=-0.015\pm 0.013$.

\begin{figure}[thb]
\centering
\includegraphics[width=0.48\textwidth]{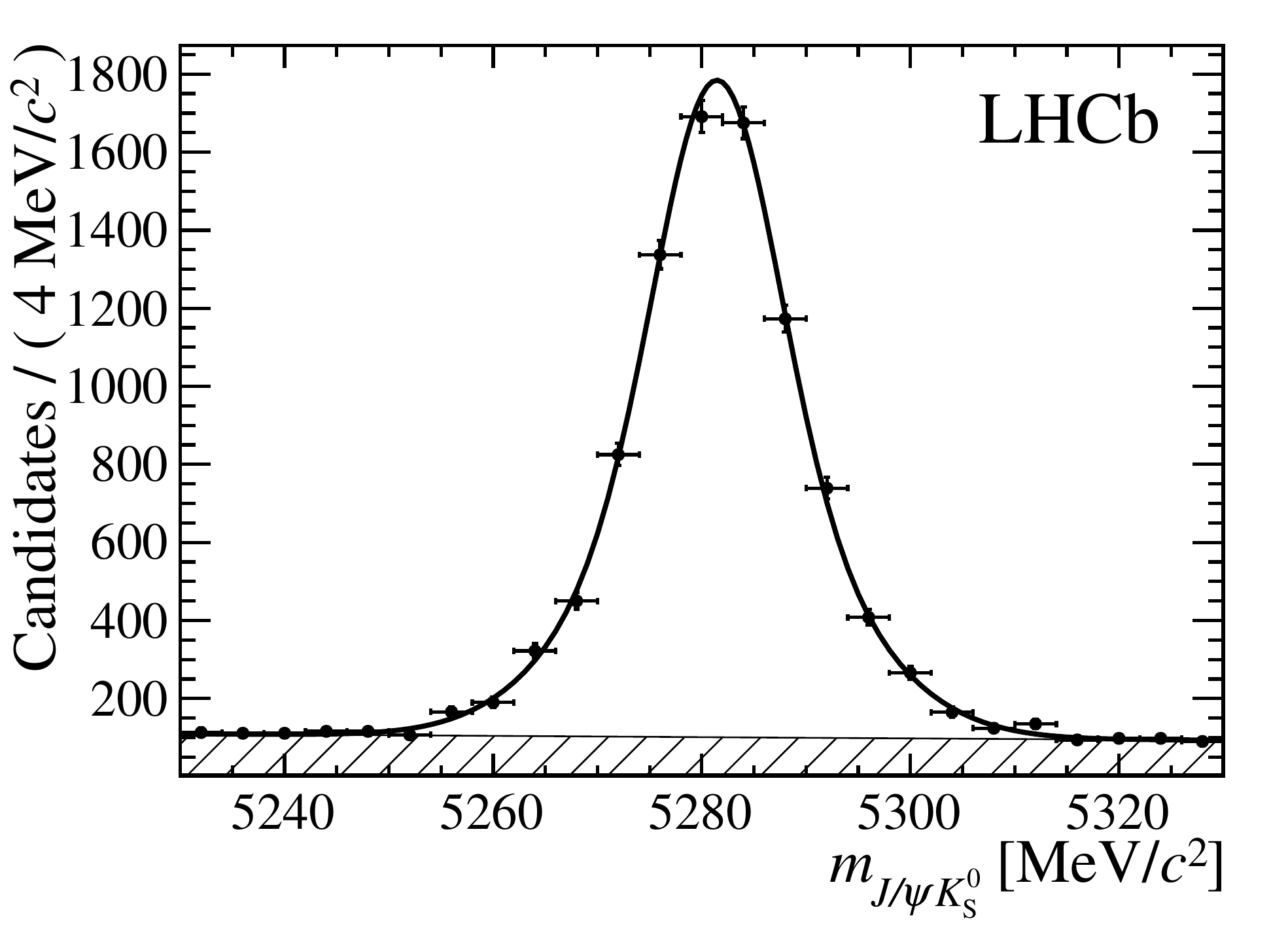} 
\includegraphics[width=0.48\textwidth]{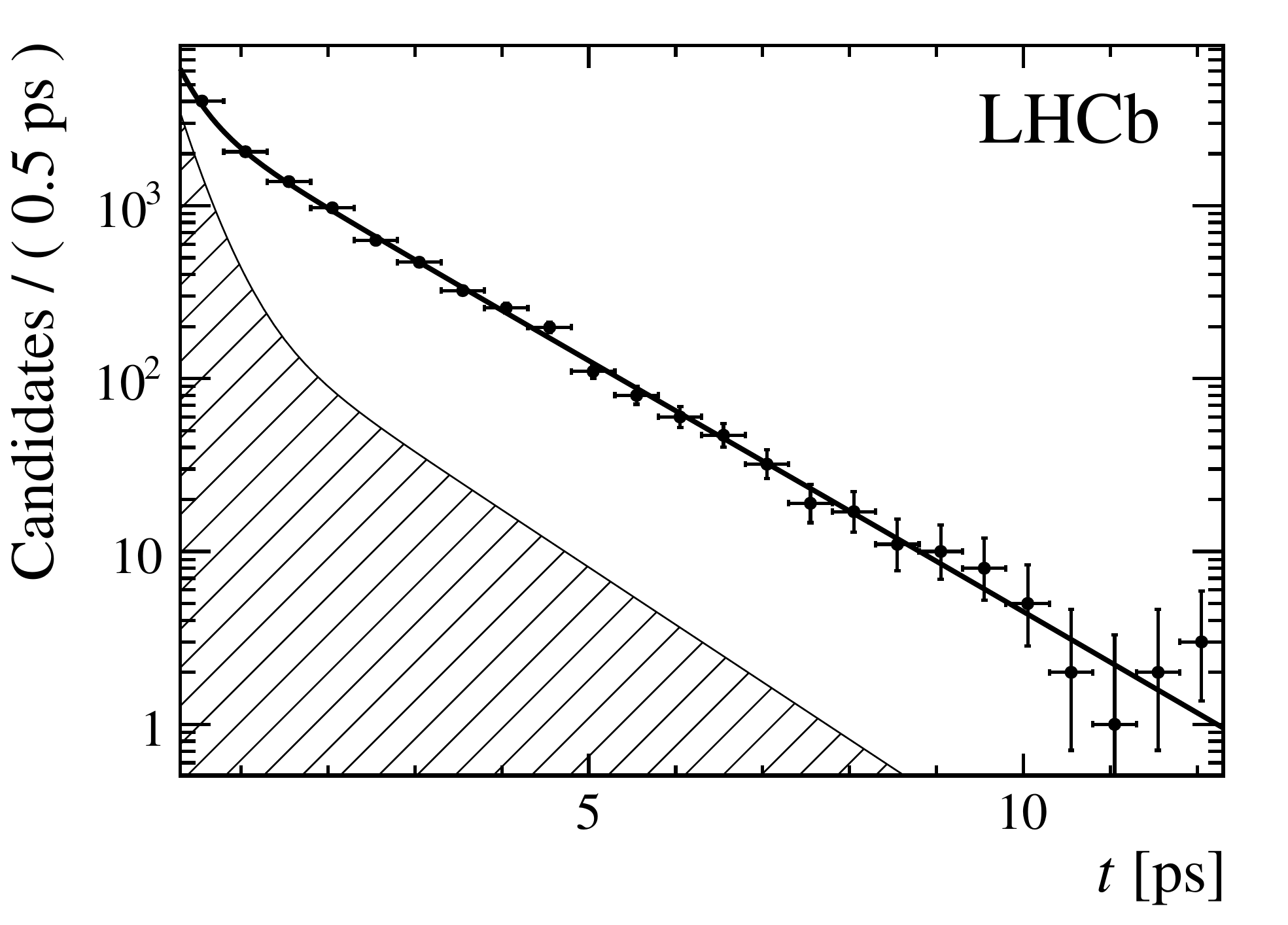}
\caption{ 
  Invariant mass (left) and decay time (right) distributions of the \BdToJpsiKS candidates. 
  The solid line shows the projection of the full PDF and the shaded area the
  projection of the background component.
}
\label{fig:plotproj}
\end{figure}

\begin{figure}[thb]
\centering
\includegraphics[width=0.7\textwidth]{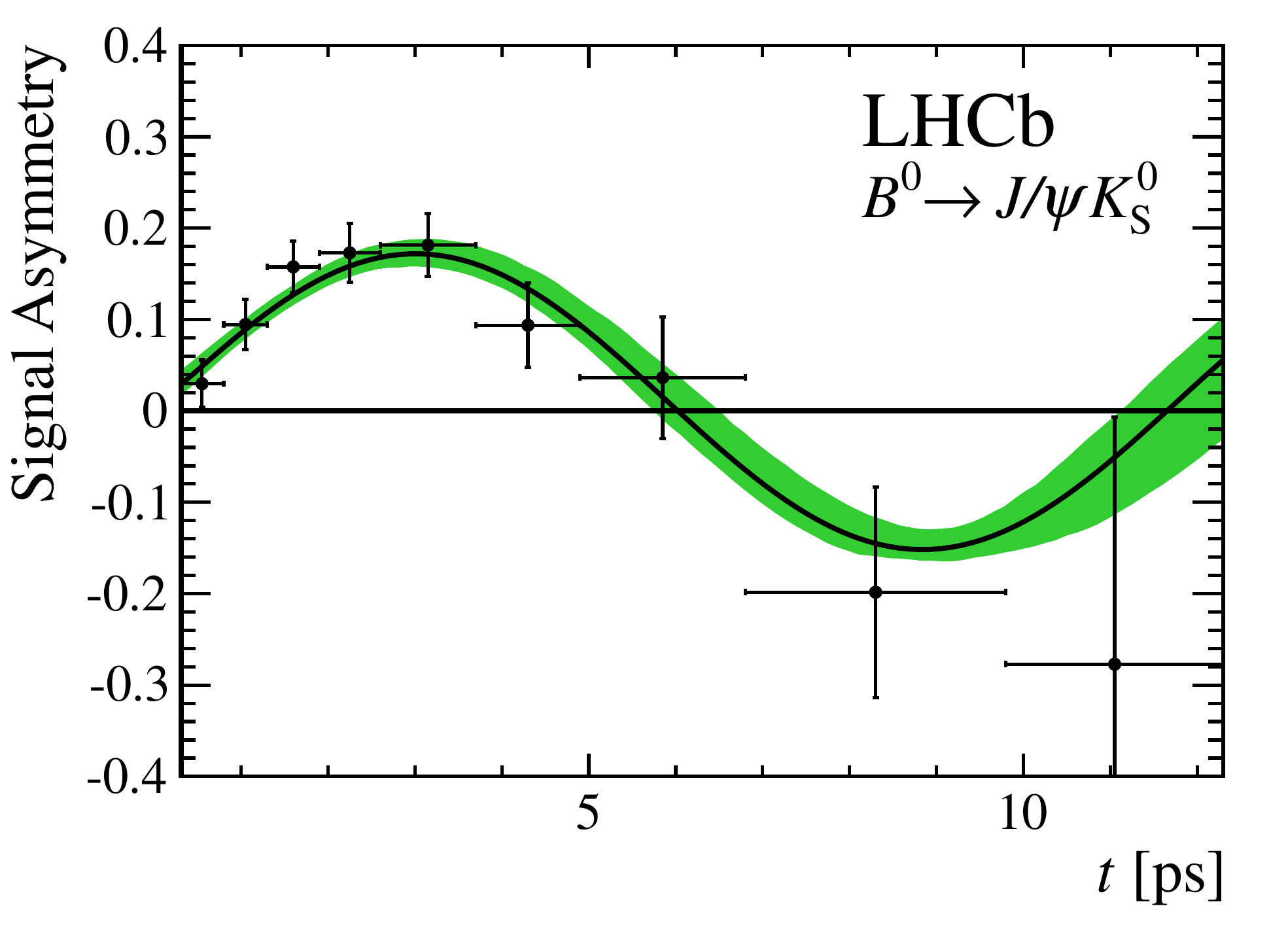} 
\caption{
  Time-dependent asymmetry $(N_{\Bzb}-N_{\Bz})/(N_{\Bzb}+N_{\Bz})$.
  Here, $N_{\Bz}$ ($N_{\Bzb}$) is the number of \BdToJpsiKS decays with a
  \Bz (\Bzb) flavour tag. The data points are obtained with the \sPlot technique,
  assigning signal weights to the events based on a fit to the reconstructed
  mass distributions. The solid curve is the signal projection of the PDF. The
  green shaded band corresponds to the one standard deviation statistical error.
}
\label{fig:asymmetry}
\end{figure}

In the fit all parameters related to decay time resolution and acceptance are
fixed. The tagging parameters and the production asymmetry parameter are
constrained within their statistical uncertainties by Gaussian constraints in
the likelihood. The fit yields 
\begin{align*}
  \SJpsiKS  = 0.73 \pm 0.07  \ ,\quad  \CJpsiKS  = 0.03 \pm 0.09  ,
\end{align*}
with a correlation coefficient $\rho(\SJpsiKS,\CJpsiKS) = 0.42$. Both of the
uncertainties and the correlation are statistical only. The lifetime is fitted
as $\tau = 1.496 \pm 0.018$\ps and the oscillation frequency as $\dmd = 0.53 \pm
0.05$\invps, both in good agreement with the world
averages~\cite{Amhis:2012bh,LHCb-PAPER-2012-032}. The mass and decay time
distributions are shown in Fig.~\ref{fig:plotproj}. The measured signal
asymmetry and the projection of the signal PDF are shown in
Fig.~\ref{fig:asymmetry}.

\section{Systematic uncertainties}
\label{sec:systematics}

Most systematic uncertainties are estimated by generating a large number of
pseudo-experiments from a modified PDF and fitting each sample with the nominal
PDF. The PDF used in the generation is chosen according to the source of
systematic uncertainty that is being investigated. The variation of the fitted
values of the \CP parameters is used to estimate systematic effects on the
measurement.

The largest systematic uncertainty arises from the limited knowledge of the
accuracy of the tagging calibration. It is estimated by varying the calibration
parameters within their systematic uncertainties in the pseudo-experiments.
Another minor systematic uncertainty related to tagging emerges from 
ignoring a possible difference of tagging efficiencies of \Bz and \Bzb.

The effect of an incorrect description of the decay time resolution model is
derived from pseudo-experiments in which the scale factors of the resolution
model are multiplied by a factor of either $0.5$ or $2$ in the generation. As
the mean decay time resolution of LHCb is much smaller than the oscillation
period of the \Bz system this variation leads only to a small systematic
uncertainty. The omission of acceptance effects for low decay times is estimated
from pseudo-experiments where the time-dependent efficiencies measured from data
are used in the generation but omitted in the fits. Additionally, a possible
inaccuracy in the description of the efficiency decrease at large decay times is
checked by varying the parameters within their errors, but is found to be
negligible.

The uncertainty induced by the limited knowledge of the background distributions
is evaluated from a fit method based on the \sPlot technique. A fit with the
PDFs for the reconstructed mass is performed to extract signal weights for the
distributions in the other observable dimensions. These weights are then used to
perform a fit with the PDF of the signal component only. The difference in fit
results is treated as an estimate of the systematic uncertainty.

To estimate the influence of possible biases in the \CP parameters
emerging from the fit method itself, the method is probed with a large
set of pseudo-experiments. Systematic uncertainties of $0.004$ for \SJpsiKS
and $0.005$ for \CJpsiKS are assigned based on the biases observed in
different fit settings.

The uncertainty on the scale of the longitudinal axis and on the scale of the
momentum~\cite{LHCb-PAPER-2011-035} sum to a total uncertainty of $<0.1\%$ on
the decay time. This has a negligible effect on the \CP parameters. Likewise,
potential biases from a non-random choice of the \Bz candidate in events with
multiple candidates are found to be negligible.

The sources of systematic effects and the resulting systematic uncertainties on
the \CP parameters are quoted in Table~\ref{tab:syst_total} where the total
systematic uncertainty is calculated by summing the individual uncertainties in
quadrature.

\begin{table}[t]
  \caption{Summary of systematic uncertainties on the \CP parameters.}
  \label{tab:syst_total}
  \centering
  \begin{tabular}{lcc}
    Origin                            & $\sigma(\SJpsiKS)$  & $\sigma(\CJpsiKS)$    \\
    \hline
    Tagging calibration               & $0.034$             & $0.001$   \\
    Tagging efficiency difference     & $0.002$             & $0.002$   \\
    Decay time resolution             & $0.001$             & $0.002$   \\
    Decay time acceptance             & $0.002$             & $0.006$   \\
    Background model                  & $0.012$             & $0.009$   \\
    Fit bias                          & $0.004$             & $0.005$   \\
    \hline                                                  
    Total                             & $0.036$             & $0.012$    \\
  \end{tabular}
\end{table}

The analysis strategy makes use of the time-integrated and time-dependent decay
rates of \BdToJpsiKS decays that are tagged as \Bz/\Bzb meson. Cross-check
analyses exploiting only the time-integrated or only the time-dependent
information show that both give results that are in good agreement and
contribute to the full analysis with comparable statistical power.

\section{Conclusion}
\label{sec:conclusion}

In a dataset of $1.0$\invfb collected with the LHCb detector, approximately
$8200$ flavour tagged decays of \BdToJpsiKS are selected to measure the \CP
observables \SJpsiKS and \CJpsiKS, which are related to the CKM angle $\beta$. A
fit to the time-dependent decay rates of \Bz and \Bzb decays yields
\begin{align*}
  \SJpsiKS &= 0.73 \pm 0.07 \text{\,(stat)} \pm 0.04 \text{\,(syst)} , \\
  \CJpsiKS &= 0.03 \pm 0.09 \text{\,(stat)} \pm 0.01 \text{\,(syst)} ,
\end{align*}
with a statistical correlation coefficient of $\rho(\SJpsiKS,\CJpsiKS) = 0.42$.
This is the first significant measurement of \CP violation in \BdToJpsiKS decays
at a hadron collider~\cite{Affolder:1999gg}. The measured values are in
agreement with previous measurements performed at the \B
factories~\cite{Aubert:2009aw,Adachi:2012et} and with the world
averages~\cite{Amhis:2012bh}.

\section*{Acknowledgements}

\noindent We express our gratitude to our colleagues in the CERN
accelerator departments for the excellent performance of the LHC. We
thank the technical and administrative staff at the LHCb
institutes. We acknowledge support from CERN and from the national
agencies: CAPES, CNPq, FAPERJ and FINEP (Brazil); NSFC (China);
CNRS/IN2P3 and Region Auvergne (France); BMBF, DFG, HGF and MPG
(Germany); SFI (Ireland); INFN (Italy); FOM and NWO (The Netherlands);
SCSR (Poland); ANCS/IFA (Romania); MinES, Rosatom, RFBR and NRC
``Kurchatov Institute'' (Russia); MinECo, XuntaGal and GENCAT (Spain);
SNSF and SER (Switzerland); NAS Ukraine (Ukraine); STFC (United
Kingdom); NSF (USA). We also acknowledge the support received from the
ERC under FP7. The Tier1 computing centres are supported by IN2P3
(France), KIT and BMBF (Germany), INFN (Italy), NWO and SURF (The
Netherlands), PIC (Spain), GridPP (United Kingdom). We are thankful
for the computing resources put at our disposal by Yandex LLC
(Russia), as well as to the communities behind the multiple open
source software packages that we depend on.

\addcontentsline{toc}{section}{References}
\bibliographystyle{LHCb}
\bibliography{LHCb-PAPER-2012-035}

\end{document}